# The Third Infoscape

## Data, Information and Knowledge in the city. New paradigms for urban interaction.

*Salvatore Iaconesi, Oriana Persico*

*ISIA Design Florence*

## Running in the city

A park, a few hundred meters away from the smells and sounds of the city. From atop the back of the gentle hills which span the green stretch of the this seventeenth century villa, urban reality seems far away, only slightly glimpsed.

Sweat, tensed muscles. The ground quickly flows below our feet, in the rapid succession of tensions and distensions of the run. That phase in which breathing breaks has arrived, abandoning the sequence of wheezy and syncopated breaths, to arrive at the condition in which breathing becomes fluid and continuous, when running becomes, in a way, automatic, free.

This modality allows for thinking, getting rid of preoccupations, fatigue, perception of effort, and to dive into the rhythm, the scanning of time and pace, as in a mantra.

And, in fact, we thing, and look around.

Far away, along the main walkway of the villa, we can see the park's café, which has just been renewed. We remember it as it was before, a residue of the 50s, with paint coming off the walls, the metal counter, the uncomfortable chairs and the small, round tables. We have never been there since it was renewed. Before that, prices were just too high and the quality too low. The only interesting thing there was Mario, who used to be an informal institution of the place: he knew every detail of the villa, and he was an mushroom expert, too, as we had the chance of discovering the many times we brought him a porcine or an honey mushroom for inspection; the ones we found close by to some tree root or in the middle of the grass fields, a little distance from there, proud to have been able to find such precious jewels freely available, instead of paying dozens of euros to purchase them at the market.

Now, Mario has retired. The management of the café has been given to a company, which has also performed the renewal. From the outside, the place looks very nice: they have augmented the glass surfaces; tables are now large and the seating more comfortable; there is free wi-fi; the industrial cornettis and the lousy orange juices have been replaced by butter croissants, organic fruit juices and classy finger food.

Or, at least, that's what we know from the many reviews we found on social networks. People have widely expressed themselves by taking carefully manipulated and filtered pictures of the dishes and of the fashionable cutlery, while lamenting the steep prices. Someone also complained about the frigidness of the service: precarious workers, led by a store manager, rapidly ending their contracts and not really getting in human contact with the place. Some of these reviewers possibly knew Mario. With him, it was not necessary to purchase anything. You could just go there, chat and, maybe, get some advice on the whereabouts of mushrooms, sage or other spices in the villa. Mario claimed that it was possible to create entire gourmet dinners for several people just by relying on the resources found in the park, including wild asparagus, mushrooms, wild vegetables, herbs, spices and fruit. You just need to know where to look for them.

One of the online comments reported how many of the employees of the new venue did not even know in which years this wonderful villa was built, or what was its purpose. The reviewer told a story in which a tourist asked for information about the park as an indolent employee answered in roman dialect «I don't know, miss! I've been working here for 10 days, and at the end of the month my contract expires. But there's free wi-fi here: if you have a computer you can look it up on Wikipedia.»[1] Mario, instead, would have taken the tourist by the arm, whether she liked it or not, and start telling, and indicating, and remembering, with the other people at the counter, waiting for their *cappuccino*.

Who knows, maybe the reviewer made the whole episode up. Or maybe not: it seemed like a sincere contribution, and not an isolated one: the same person demonstrated a systematic effort in reviewing multiple places across the city, among bars, restaurants, museums and art galleries. Eventually, we will go to the new bar in the villa. And, maybe, we will also review it online, including the story of Mario and of his knowledge of mushrooms. Or, maybe, we will keep that part of the story for ourselves, to avoid hordes of improvised mushroom gatherers to start collecting anything that even vaguely resembles a mushroom, or even that someone might get sick with them: not everyone is as good as Mario in recognizing their edibility. Given this, it would be nice if all of Mario's knowledge could, somehow, have remained in the park, accessible and available. Someone could have made an *app* for it: even Mario himself, to continue to be present, with all of his knowledge, in the seventeenth century villa which has welcomed him for all of these years.

Here, a steep climb is next. Beyond that is the valley where the small lake at the center of the villa lays, rich in ducks, swans, crows, turtles and many other animal species.

Heart beats mad, we're breathless again. It's been a couple of weeks since we have last been running.

We check our smartphone.  A map shows the path of the run. An information visualization indicates the state of our heartbeats and breathing, as captured thanks to the biometric sensor positioned into our watch, and shows us the comparison with the trends from last weeks' runs, together with counts of burned calories and the juxtaposition with our friends' performances (oh, there is one of our friends in this same part, right now, we notice, right behind the next hill... maybe we can catch up with him).

---

[1] «Ecchenesò signò! Io lavoro qui da 10 giorni, e a fine mese mi scade il contratto. Ma qui c'è il wifi: se c'ha un computer lo può cercare su Wikipedia.»

Climb. Heartbeat. Breath.

A notification arrives.

It's our partner.

The message is funny and goliardic: «Put on some weight? A slight hill and you already can't take it anymore? ;)»

We're not used to it, yet. The app was a peculiar anniversary gift, together with the smart-watch to which it is connected. Real-time heartbeats, visualized on your partner's smartphone. It looked like a romantic idea and, in practice, you don't even remember about it, in your daily routine. But when notifications arrive, as in this case, a weird sensation emerges, ranging from strange sinister to magical.

The map alerts us: our friend is right behind the turn. We have almost caught up with him. We step beside him and exchange a quick chat while running. We cheer each other up, and decide to embrace the climb, which will take us at the top of the highest hill in the park, where we will enjoy a beautiful view of the city.

Reaching the top, we pause to look at the scenery, and to stretch a bit. Then we separate: we have our car on the other side of the park, our friend has it on the exact opposite side, and it's getting time to go back home.

The descent is exhilarating: trees flow by fast on our sides, while our legs almost fail to keep up the pace with the sloping ground, sliding like a crazed tapis-roulant below us. We slip into a green tunnel, a spot where trees and plants interlocked themselves from both sides, forming a complex architecture above us.

The declivity is almost over when our smartphone, fixed with a strap to our arm vibrates again: another notification.

We slow down and reduce the pace, obtaining a better balance which allows us to check the origin of the notification: a message from the office; it's urgent.

We stop, take breath, and call, with the small towel on our neck avoiding us catching a cold through the cooling sweat.

When they answer, we're still partially breathless: «What's going on?»

An agitated dialogue: there is a problem on a project, and fast decisions must be made. We find ourselves in a conference call with all the rest of the office: connected simultaneously from three cities and, from what it looks like, from only one park. During the discussion the park disappears. For those few moments we feel in another, different place. Only our shorts and t-shirt there to remind us that we've been running in the seventeenth century villa, amidst nature, which also remains at the border of our perception. We are in another place now, somewhere between/across the multiple places of the conference call. We're, momentarily, in an office, not in a villa, that's for sure.

An ubiquitous office.

The conference call ends, and we start running again and, slowly, the park starts reappearing. While our running commences once again, we manage to progressively abandon our ubiquitous office, with only the last few thoughts diverting our attention from the nature around ourselves, the trees, plants, sounds and shapes of the woods and hills, which powerfully claim back our consciousness.

We arrive at the gate of the villa. Our car is parked just outside. We check our smart-watch: twelve kilometers. Not bad. We touch a dial on the interface. Done. Our run gets automatically published on social networks, together with maps, times and calories, ready to be shared and commented.

## Reading the city

There are many informational elements which we experience when we interact with cities.

Traditionally, they were physical: architectures, visuals, signage, sounds, smells.

In his *Concise Townscape*[2], Gordon Cullen provided a seminal description of the modalities which can be used to plan and design the shape of the city in order to stimulate emotional and interactive responses. The book emphasizes the role of personal experience in urban landscapes. This concept is remarked by highlighting the many ways in which we actually *create places,* under the form of mind maps: *serial vision, juxtaposition* and *immediateness*. The analysis of the visual elements which allow us to comprehend public space and to understand its usages and rules brings us to the exploration in which city dwellers spatially perceive information in the continuous dialogue between the actual visual composition of the environment and the one which emerges as they traverse the city. In this kind of analysis, juxtaposition and immediateness (then declined as legibility or accessibility) of the elements of spatial information become emotional and operational actuators of the interactions between the city and its users[3].

Kevin Lynch, in T*he Image of the City*[4]*,* started from a few case studies to explore the legibility of the city, formulating the hypothesis according to which people construct mental maps used to traverse, interact and relate with urban spaces and the population. The possibility to collect and compare these mental maps enables to comprehend how visual, perceptive, cognitive and psychological information in urban territories contribute to forming models of people's interactions and relations.

These and other texts – and in the practices connected to them – tend to account for physical elements of reality, and include the symbolic and cultural dimensions.

In *Learning from Las Vegas*[5] Robert Venturi, moved beyond that. By observing the surreal and suggestive visions of the city of casinos in Nevada, he built an image of cities as expression of a

---

society which progresses towards a reality constructed not only through symbols, but also through reproductions and representations.

This information comes under the form of iconographic instances, and constitute a series of stratifications – of layers – on the territory: an immaterial geography which contributes to the modeling of citizens' mental maps and, thus, to the ways in which they act and react, just as the physical shapes, the sounds, smells and tactilities.

Aldo Rossi, in conflict with Venturi's *Complexity and Contraddiction in Architecture*[6], merged a value and time based analysis to the symbolic and post-modern perspective, alluding to the possibility of imagining the city as something which could be constructed with the flowing of time and the superimposition of the stories of its inhabitants, referring to the opportunity to build the place of the dynamic preservation of the collective memory[7].

These considerations may be added to the ones by Lefebvre[8], De Certeau[9] and Soja[10], among others, according to whom the construction of space and, thus, its communication and interaction, is composed by a performative aspect which is polyphonic, and in which the *micro-histories* of people (according to the definitions by Levi[11] e di Iggers[12]) dynamically and generatively interweave in the continuous recomposition of urban space.

In *Walking in the City*, De Certeau writes:

«The ordinary practitioners of the city live "down below", below the thresholds at which visibility begins. They walk – an elementary form of this experience of the city; They are walkers, Wandersmänner, whose bodies follow the thicks and thins of an urban "text" they write without being able to read it. These practitioners make use of spaces that cannot be seen; their knowledge of them is as blind as that of lovers in each other's arms. The paths that correspond in this interweaving, unrecognized poems in which each body is an element signed by many others, elude legibility. It is as though the practices organizing a bustling city were characterized by their blindness. The networks of these moving, intersecting writings compose a manifold story that has neither author nor spectator, shaped out of fragments of trajectories and alterations of spaces: in relation to representations, it remains daily and indefinitely Other.»

Then continues:

«Their story begins on ground level, with footsteps. They are myriad, but do not compose a series. They cannot be counted because each unit has a qualitative character: a style of tactile apprehension and kinesthetic appropriation. Their swarming mass is an innumerable collection of singularities. Their interwined paths give their shape to spaces. They weave places together. In that respect, pedestrian movements form one of those "real systems whose existence in fact makes up the city".

---

6  Venturi, Robert. Complexity and Contradiction in Architecture. 2nd edition. New York: The Museum of Modern Art, 2002.
7  Rossi, Aldo. L'Architettura Della Città. Padova: Marsilio, 1966.
8  Lefebvre, Henri. The Production of Space. Hoboken, NJ, USA: Wiley-Blackwell, 1991.
9  De Certeau, Michel. The Practice of Everyday Life. Berkeley: University of California Press, 1984.
10 Soja, Edward W. Thirdspace. Malden: Blackwell, 1996.
11 Levi, Giovanni. "On Microhistory." In New Perspectives on Historical Writing, edited by Peter Burke. University Park: Pennsylvanya State Press, 1991.
12 Iggers, George. "From Macro to Microhistory: The History of Everyday Life." In Historiography of the 20th Century. Hanover: Wesleyan University Press, 1997.

They are not localized; it is rather that they spatialize. They are no more inserted whithin a container than those Chinese characters speakers sketch out on their hands with their fingertips.»

According to Soja this modality constitutes a Third Space in which

« everything comes together… subjectivity and objectivity, the abstract and the concrete, the real and the imagined, the knowable and the unimaginable, the repetitive and the differential, structure and agency, mind and body, consciousness and the unconscious, the disciplined and the transdisciplinary, everyday life and unending history.»

The Third Space is a radically inclusive concept, in which strategies and tactics coexist polyphonically and, thus, where the latter achieve visibility and legibility, allowing for the liberation, contestation and re-negotiation of boundaries and of cultural identities.

## Ubiquitous Information

Data, information and knowledge are ubiquitous.

They are in the shapes of buildings, in streets and urban furniture. In the forms chosen by city dwellers to traverse spaces and places. In signs, symbols, images and icons. In colors. In the smells and sound we feel while we're in the city. In the skyline. In objects which are near, and in those we see at the horizon. In the memories which we associate to places, objects and contexts, and in those memories which other people described to us, as we remember them, in precise ways, or not.

All of these – and more – elements – whether they are physical, immaterial, ephemeral –, all of these sources of data, information and knowledge, contribute to the shapes of our actions, to our performance of the city.

Ubiquitous technologies and digital networks largely augment our possibility to generate data, information and knowledge and to experience them, everywhere and at all times. And, thus, to shape our urban performance.

Smartphones, sensors, services of all kinds, interactive screens, urban screens, media facades, augmented realities, social networks, natural interaction systems activated by gesture, voice and movement. These and other devices, interfaces and services add up to other one – already more common and diffused – wich allow us to transform each action in data and information, wheter we realize it or not, consciously or unconsciously. It is the flipping on or off of a light switch; the passage within the field of view of a security camera; the querying for some train information using an interactive billboard; the purchase using a credit card; the usage of an app on our smartphone to find a restaurant we like; the notification on the health condition of one of our loved ones, generated through a biometric sensor; or the many other available possibilities through which any gesture action can be transformed into data and information.

The life of these data and information is complex.

Some of them are archived, according to various modalities and degrees of persistence, until they are needed.

Other are directly used, to execute services and processes.

For example, if person A desired to execute a purchase using a payment system on their mobile phone (maybe using *Bitcoins*[13]) a few clicks or touches on the device would be sufficient to execute the necessary data checks to authorize the transaction (generating more data).

Even in this apparently simple case, many interesting questions could arise.

For example, a friend of A could find himself in the shop and, realizing that he left home without cash or credit cards, could have asked A to execute the purchase for him, promising to give back as soon as possible. The A subject could be thousands of miles away from the store and, nonetheless, execute a physical purchase in that location.

Other modalities are possible, ranging from the more simple and *legible* ones (such as swiping an RFID[14] card in the immediate proximity of a reader to access the subway or to pass a toll booth), to the more complex and *opaque*.

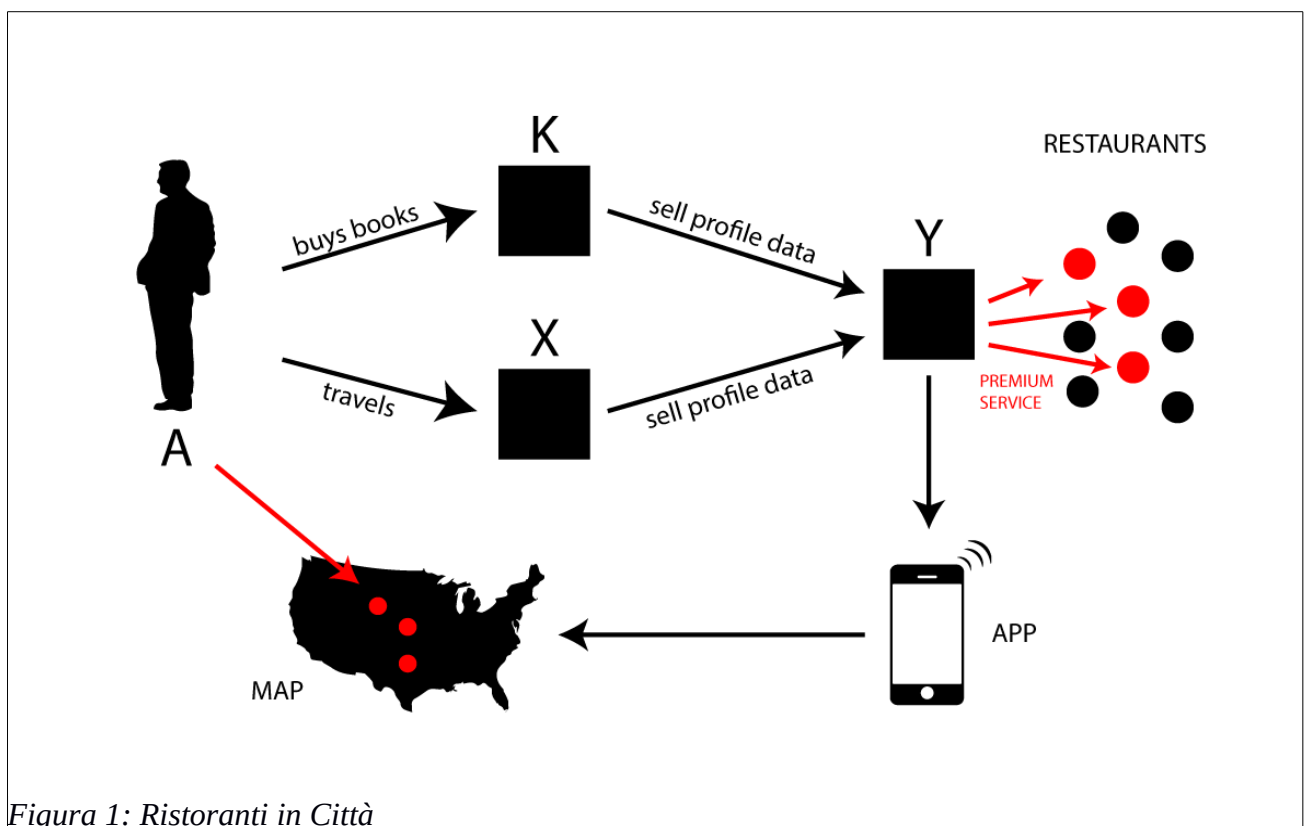

*Figura 1: Ristoranti in Città*

We could choose to analyze these last ones.

As shown in figure 1, the A subject is a constant visitor to website K, which he uses to purchase books and other objects. The website records A's preferences and tastes, by observing the characteristics of the product pages he visits online, his purchases and other parameters (for example the time A spends on each product's page). The service is able to build a profile for A whose quality is proportional to the quantity of time and interactions spent on the website.

---

13  Bitcoin is a digital currency created in 2009, which allows to execute anonymous economic transactions. For more information it is possible to look at: http://it.wikipedia.org/wiki/Bitcoin
14  RFID means Radio Frequency IDentification, and indicates a series of technologies which can mark objects with tags which can be read and updated automatically by dedicated readers in their immediate vicinity. The tags often come under the form of small stickers containing a passive radio frequency circuit. For more information it is possible to look at: http://it.wikipedia.org/wiki/Radio-frequency_identification

Another service, X, executes similar profiling processes, for travel and tourism.

To access services K and X, subject A has accepted their terms of services[15] of both websites, which attest that the profiling information can be shared with third parties.

A certain company, Y, is one of these third parties. Both K and X share (sell) profiling information to Y. Y integrates the data, obtaining new data, maybe discovering that subject A loves oriental cultures, and that he frequently visits Japanese restaurants.

Y, maybe, manages a social media site dedicated to food and restaurants, and massively uses this kind of data to decide its advertising strategies, providing its advertisers (restaurants, which probably advertise their commercial venue using Y) a premium service in which they are granted that their ads will be highlighted for the users whose profiles clearly indicate that they have a preference for the type of food or style of the advertiser's venue, and that, thus, they may have a higher probability of ending up actually eating at their restaurant.

Y also has a smartphone application, installed on subject A's device.

A finds himself in a new city, which he knows little about. Dinner time comes up, and A wishes to select a restaurant. To do it, he uses application Y. On his mobile phone's screen he will see a map, on which the Japanese restaurants which have purchased Y's premium service are highlighted and, with all probability, will choose one of them to spend his evening.

This is a very complex (and very frequent) modality for interaction. For example we can imagine replacing X with Amazon[16], K with TripAdvisor[17] and Y with Yelp[18].

A user's interaction with social networks and online websites generates data and information which is algorithmically recombined in complex ways and used by other services and websites which (in this case) generate maps and lists of commercial services in personalized ways which are completely opaque to users.

The user has no way in which he can learn which information he produced have been used as a source of information to generate the map in the Y application. Furthermore, these sources may vary each day, according to the evolution of the commercial agreements running between companies.

The result is that subject A sees a map which provides a complex vision of the reality of his surrounding territory, in which the highlighting of the restaurants is the result of a complex elaboration of his tastes and desires, and of how X's and K's algorithms (both invisible for A) have been able to identify them.

The map provided by Y represents a complex geography. It is, in part, physical (it effectively represents the position of buildings, streets and restaurants), emotional and psychological (it derives from the interpretation of A's tastes), partly economic and financial (being determined by the fact that some restaurants have purchased Y's premium service or not), and we could go on in detailing the list of its technological, algorithmic, statistical, anthropological characters, and more.

---

15  Terms of Service Agreements are legal documents which users of online systems accept when using certain websites or services, and which regulate their characteristics, also in relation to the existing laws and regulations. The acceptance may be explicit (for example when users are forced to click on an "Accept" button as they are subscribing to the service) or implicit (for example when certain websites display messages like "By using this website the user declares that he/she has viewed and agreed to the Terms of Service agreement available at [link].")
16  Amazon: http://www.amazon.com/
17  TripAdvisor: http://www.tripadvisor.com/
18  Yelp: http://www.yelp.com/

This complex scenario generates a multitude of different ways in which these types of information can be transformed into opportunities for interaction with citizens and other human beings.

Furthermore, the information can be recombined in infinite different ways, to produce infinite different geographies, opportunities for interaction, services, games, artworks and more.

We find ourselves well beyond Mitchell's *City of Bits*[19], or of Zook's and Graham's description of *DigiPlace*[20]. Just as we're well beyond the idea of geo-referenced data backing these two visions, and also of the ones expressed by Aurigi[21], and in their articulation of space in its digital and physical character.

We find ourselves in a digital Third Space, more inclusive, in which information is not only attached to places, spaces, bodies and objects, but constantly recombines, remixes, recontextualizes, creating constantly new geographies which are emotional, linguistic, semantic, relational, or relative to the many patterns which non-human algorithms can glimpse in the ways in which layers emerge from data, information and knowledge, correlating different spaces, times and human networks.

## The Third Infoscape

Gilles Clément describes the *Third Landscape*[22] as an *uncoded* space, the space of biodiversity which is able to host the genetic reservoir of the planet. The Planetary Garden it a space for the future, for the emergence of possibility[23]. It is also a connective tissue composed by the unison of residual spaces which assume fluid forms, which are able to escape form and governance. They are places which cannot be preserved through administrative dimensions, which would destroy their characteristics. Barrell's *Dark Side of the Landscape*[24] comes up to mind, and his description of the ways in which the natural landscape of cities derives from the imposition of the point of view of a single social class. Clément, instead, speaks about a *light side*, as the Third Landscape does not represent an exclusive model, but an inclusive one, a shared fragment of a collective consciousness. It is a multiplication of narratives, a planetary remix (*brassage*) in which perennial mutating spaces incorporate the presence of multiple representations: syncretic maps which describe the geographies of the mutation of the city.

Clément also tells us about the need to educate our gaze to recognize the Third Landscape, to recognize emergence and to transform it into shared knowledge.

---

This is similar to the concept of *ruin* expressed by Marco Casagrande.

A ruin represents the progressive reunion of objects and architectures with nature: nature and human beings *ruin* buildings and objects, transforming them into ruins.

From a different point of view, these actions bring objects and buildings in a different state. A ruin is also the evidence of the history of human and natural action, of the daily usage *patterns*. From this point of view ruins expose everyday life, in all of its complex manifestations.

Therefore, ruins can be considered as the progressive layering of stories, as a source of information and knowledge.

Casagrande uses the concept of the ruin to define the Third Generation City as the «ruin of the industrial city»[25] and as the «industrial city ruined by people – human nature as part of nature.»[26]

The concept of Open Source infiltrates in the text:

«Like a weed creeping into an air-conditioning machine the industrial city will be ruined by rumors and by stories. The common subconscious will surface to the street level and architecture will start constructing for the stories - for the urban narrative. This will be soft, organic and as an open source based media, the copyrights will be violated. The author will no longer be an architect or an urban planner, but somehow a bigger mind of people. In this sense the architects will be like design shamans merely interpreting what the bigger nature of the shared mind is transmitting.»

In this vision the city assumes the shape of a body in perennial mutation, including both architectures and the constant and emergent layering of stories and knowledge which originate from the daily lives of citizens and nature.

At this point, it is possible to go back to our original narrative, to the concept of ubiquitous and emergent knowledge, and to connect it to this vision, to Clément's and Casagrande's vision.

The new types of information, the ones which converge in our perception of the city and, thus, into our interaction with human beings, architectures, spaces, places and organizations, be them emotional, semantic, linguistic, relational, relative to the possibility to identify *multi-modal* and *multi-layered patterns* which can be localized anywhere in space-time, whether they are generative or algorithmic, whether they derive from sensors or other interactions... all of these instances of data, information and knowledge, today, often have digital form and ubiquitous manifestations.

We experience them through smartphones, applications, social networks, interactive services and systems which are disseminated, distributed or even pulverized through space and time. Through them we can interact with the world, express ourselves, collaborate, work, express emotions, consume, study, entertain ourselves.

Following the previous examples it is, thus, possible to attempt the definition of the Third Informational Landscape: the *Third Infoscape*.

The First Infoscape refers to the information and knowledge generated through the modalities of the

---

pre-industrial city. The Second Infoscape refers to the information and knowledge generated in the industrial city (the Second Generation City, the city of infrastructures, transactions, sensors...).

*The Thirs Infoscape refers to the information and knowledge generated through the myriads of micro-histories, through the progressive, emergent and polyphonic sedimentation of the expressions of the daily lives of city dwellers.*

The vision of the new paradigms of interaction with the city are centered on the Third Infoscape.

**A walk in a new city**

*Empathy*.

I still did not get used to this new App. It is facinating and disorienting, at the same time.

It is a form of telepathi. That's the most effective way in which I can describe it. Telepathy with things, people, environments and systems of any kind.

I am here, in this bar, anonymous, waiting, sitting at a small table.

As soon as I arrived a waiter, after freezing for a bit – assuming the typical absent expression of when you are accessing *Empathy* –, smiled and started talking to me: «There is a Japanese *noise* music concert on Friday.»

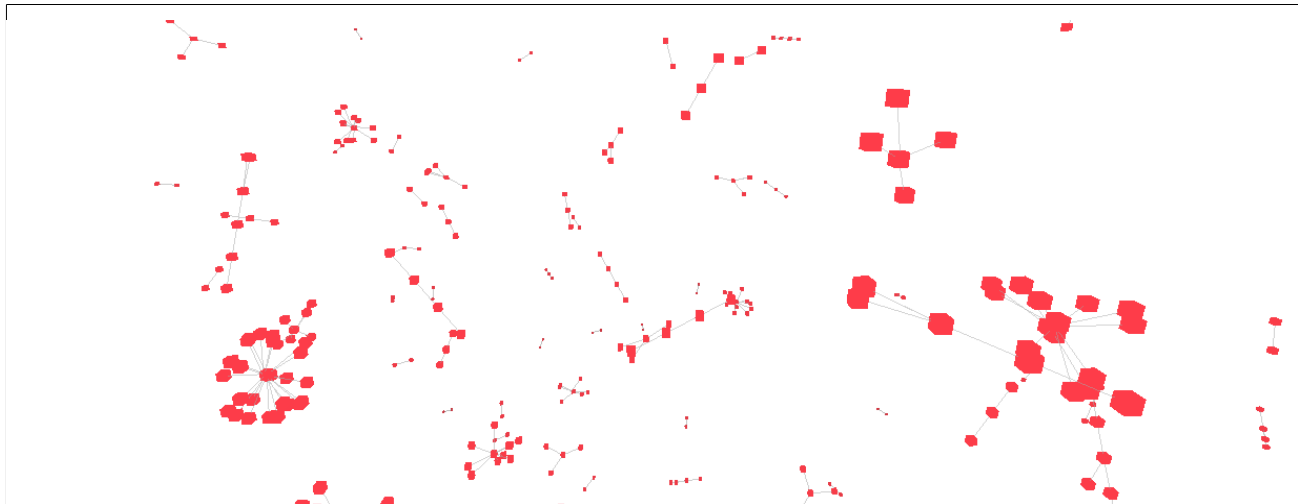
*Figura 1: Una piccola parte del Grafo Relazionale Pubblico della città di New Haven*

In the *public relational graph* of the city, I am one of the main nodes for topics related to music of this kind. I can still remember the time when, at a business meeting, the clients started to giggle: «Do you *really* like that crap music?»

Maybe I should reconsider what to express publicly on *Empathy*.

Positioning myself on a series of interests, topics and flows of information seemed like an intelligent thing to do, at first. But the more I go on, the more I understand how it could be better to select and filter, and to construct my *personal relational graph* more slowly, with care, only after having understood people and organizations more profoundly.

Now, for example, I really don't want to talk with this guy. Now that we interacted and, thus, that we're connected, I can perceive his positioning on the *network*.

While I wait, with a smoking tea in my hands, I *empathize with the place*.

*I perceive a peak of joy in German, on Wednesday afternoons*. I investigate: according to social networks every week, on that day, a community of Germans meets here. It is a recurring pattern, and they expose it in public to enlarge the community, to be able to have a moment in the week to be able to speak with people of the same nationality, here, so far away from home.

While I analyze the other *trends* in this place (but, where is my appointment?), I loose myself a bit in the ubiquitous empathy.

I navigate *filters* and *perspectives*.

I get to know that temperature has lowered, and that the level of fine dusts suspended in the air of this neighborhood is very high (just perfect for my soar throat). I know that a just a few blocks away a citizen meeting has just taken place, to decide the placement of an urban vegetable garden in that park, two further blocks away, which has been at the center of recent critiques for its state of abandon.

I visualize their human network, and I see it energetic and healthy, with emotions, opinions, information and knowledge continuously flowing from one node to the other. They are understanding how to implement the project: someone is analysing administrative requirements; someone the technical specifications (irrigation, tools, similar experiences...); someone figuring out how the production of the garden, with its fruit and vegetables, would be split among citizens; someone figuring out the economic aspects of the operation. There is also a raging discussion going on about the fact that the sudden free availability of food in the garden could cause the presence in the neighborhood of undesirable people – that's the term they use most, according to the *natural language analysis* algorithms –, like clochards, homeless, people of the street.

A city councilor also entered the discussion, ready to support the initiative and to measure its effects (always through *Empathy*) in terms of social relations, inclusiveness, accessibility and emotions (does this type of initiative generate more well-being, anxiety, stress, or what?).

While I wait, I share a web link with one of their influencers, about an existing experience about urban edible woods: beautiful place to see, used to regenerate interstitial and marginal areas of the city, and which produce vegetables, fruit and citizen social activation. They create communities and feed people.

I observe my watch. The time of my appointment has long passed. I finish up my tea, pay, get up and leave.

I consult the map and the calendar. I could go to another meeting using the subway, or go back home. I use *Empathy* to take a look at the urban *dashboard*. Along the subway line, in the part of the journey which I would be on to go to the meeting, lots of people are expressing stress and indignation: for some reason trains have slowed down and, thus, wagons fill up excessively, with people crushed and angry. Traffic is not reassuring, either.

I decide to walk.

I go up to the street light, to cross to the other side of the road, while I keep looking around.

The sign of a Korean restaurant catches my eye: it is composed by an English writing ("The Red

Dragon") and from one written in Hangugŏ (나머지 산, which roughly translates into "The Refuge atop the Mountain", as I see from the augmented reality translation which appears superimposed to it, in light transparency). Hundreds of people, before me, noticed it, as I can see through *Empathy*, with the profusion of photos and comments published by people of all kinds on social networks. I join the crowd: «How comes that "Red Dragon" translates into "Refuge atop the Mountain"?» A user clears up the issue: «It is a cultural reference: traditionally, many inns were found atop mountains, in which you would go to find tranquillity, good food and liquors. It has become a figured speech which is hard to render properly in English. That restaurant is very popular with Korean citizens who go there to...»

While I read the reply (interesting!), still asking myself how this reference translates into "Red Dragon", *Empathy* does not rest, and continues proposing information to my attention.

Nearby is a bookstore which has in stock some publications which are in my Amazon wish-list.

Images from my apartment, where the plumber has just finished fixing a leak. Together with this, a visualization of the energetic efficiency and consumption levels of my home, raising to a better level.

A report from my office, together with the visualization of the emotional expression of my colleagues, progressively becoming more nervous for the arrival of a tough deadline.

One of the notifications is really fun: our block of homes, thanks to the leak I just fixed, has become the most hydrically efficient one in the *Game of Energy in the City*. On *Empathy* the entire block will shine brightly, literally.

I decide to cut through a park, within a beautiful seventeenth century villa, in the heart of the city. I remember when I used to go running here, a few years ago: the bar in the middle of the park had just been renewed from the previous management, of Mr. … how was he called... Mario. Mario, who knew so much about the villa, and abut mushrooms: you could feel his absence when he retired.

I start thinking about what Mario would have thought about *Empathy*, wether he would have considered a weird technology, or as an opportunity.

While I wander along the parkway, moving onto the grass, setting about to traverse the small wood which separates me from the other side of the villa, closer home, I decide to turn on the *Third Infoscape*, one of the most controversial extensions of *Empathy*.

On *Third Infoscape* there are no barriers of limitations: all the data, information and knowledge produced by services, people, sensors, algorithms and by anything which is able to generate data, information or knowledge, are available in real-time (almost... but they are making the system better), and they are accessible for you to *remix*, recombine, correlate with all the rest.

This modality always reminds me of the ways in which plants grow: it is like the grass that grows between the cracks in the wall, or amidst the train tracks, or wherever it is that it finds some space. It grows. Full stop.

Just like the *Third Infoscape*. It is like seeing data growing, and information, and knowledge, in the informational, relational, emotional and knowledge ecosystem in the city.

It is really interesting to see, just as interesting as when go to cities like Sao Paulo in Brazil and you see urban nature – pigeonholed within flowed beds, gardens, boundaries and confines – waiting, energetic like a compressed spring, for an opportunity to express, even for a day or two. When this happens, growth is immediate: the bricks surrounding the flower beds break; the concrete borders are invaded by roots and leaves; pavements grow holes in which new growths form.

*Third Infoscape* is like this: data, information and knowledge growing everywhere, just like in a crazed Sao Paulo, in which all the gardeners of the city administration simultaneously fired themselves.

It is a beautiful thing to see, as nature growing: data emerging from places, people, objects, spaces, buildings, everything. As roots, leaves, extensions, connections.

It seemed beautiful but useless, initially. They seemed like too much data, all together. But, progressively, people have started to understand it, and to use it. And, in doing that, they reminded me of Mario.

I turn on the *EatNature* filter on the Third Infoscape, collaboratively created by people from all over the world to highlight the parts of the natural environment (woods, bushes, shrubs, herbs...) which can actually be eaten, and which has multiple manifestations here, in the villa. For example, walking in this small wood where I am now, I could add this walnut tree to the filter, or, on this other side, this small bush of wild rocket salad.

People have progressively learned how to use it, and to bring up new types of economies. In ways which are similar to what happens in the physical environment. There is wood? I can build furniture. Marble? Produce statues. Artisans and merchants? Bring up a district.

These other materials are aboundant, infinite, recombinant. You don't need to dig, cut or kill to use them. They do not consume or expire. They are the data coming from sensors, transactions, relations, emotions, opinions, movements, desires, people, non-people.

While I stroll, I am surrounded by trees, data, information and knowledge.

I am wearing my headphones, in which a generative music runs. It has been created by an artist who uses the data which is able to express the well-being (or not) of the natural environment, as it is captured through sensors and people observation, to create a continuous, infinite symphony: I can listen to the happiness (or sadness) of the natural environment around me. I see the names of the plants, in augmented reality; the places where people have been happiest, melancholic or amazed.

Mario would have plenty of information and knowledge to contribute.

I leave the small wood, and see the park's exit.

I cross it, and find myself on the street, again.

To my left, a beautiful ex-industrial building. It seems as if it has been abandoned since a long time.

Using *Empathy* and *Third Infoscape* I learn that it is a building owned by the city administration, unused and decommissioned from multiple years. There are many citizens wodering what to do with it. The administration is using *Third Infoscape* to make people come together, creating relations and opportunities to understand what to do with it, by activating which citizens, on which themes and through which forms of economic sustainability.

I have an idea, and I publish it, in-between the network of proposals: a place which is a mix of an university, an agrarian research center and a citizen laboratory, in which the villa is used to discover the natural environment, food, how to produce it, using ancient seeds, to prouce it in sustainable and autonomous ways, to study, cook and eat together. Almost immediately, someone replies: they are two citizens and a nearby restaurant, whose owned is passionate about the issue of the ancient seeds and of biodynamic agriculture.

He would like to join in the idea.

I'm in the network.

## A scenario for future cities

Graham[27] wondered how it could be possible to imagine a real time city by taking in consideration the ways in which telecommunications reconfigure our notions of time and urban space. This goes in the direction of the definition of a communicational environment, a diffused *cloud* of sense and meaning which goes beyond the dynamics of screens, and which is not virtual anymore, but impalpable and mental.

This atmosphere is found in the spaces which are *in-between*, interstitial, *ubiquitous*. It is not an idealized representation, but a mobilization of imperceptible urban matter, manifesting itself through pervasive computation which is both automatic and relational.

To all effects, with the development of wireless sensors, of *smart dust*[28], and with the possibility to engage human beings in urban sensing processes, the dimension of virtuality collapses. Heading towards a state which is basically comparable to the one of *telepathy* (among human beings, human beings and machines, machines and machines...), reconfiguring urban ecologies so that mapping virtuality or physicality would not be needed anymore, and replacing this need with the possibility to create recombinant inventories of the telepathic migration of dusts, of the myriads of pulverized sensors which are disseminated, diffused.

This *telepathic* form is, thus, a form of invisible communication which describes the ways in which the city talks to itself, circulating messages and reprogramming urban ecologies.

The circulation of messages represents and moves physical shifts and transformations. The city itself moves, as phenomenon and meta-phenomenon. A feedback loop, thus, is created, in which we find ourselves simultaneously immersed and unaware of the – telepathic – exchanges which surround us.

We can imagine information mutating into landscape, delineating an urban space which is not determined by distance and time, but from the transformation of densities and presences.

Gabrys states that:

---

27   Graham, Stephen. "Cities in the Real-Time Age: The Paradigm Challenge of Telecommunications to the Conception and Planning or Urban Space." Environment and Planning A 29, no. 1 (1997): 105–27.
28   Le smart dust sono nuvole di microscopici sensori, collegati da reti wireless, che possono essere diffusi nell'ambiente. Per saperne di più è possibile consultare: http://it.wikipedia.org/wiki/Smartdust

«The wireless city is a space for the production of dust in all its modalities. The city abounds with compressed and errant signals. Yet instead of dissolving urban space, as so many writers suggest, these communication and sensing technologies fill it with signals.»[29]

It is interesting to note, after all, how it is not important that messages arrive to destination and accurately assemble themselves, but that it possible to understand how these are filtered by *noise* and *dust*, and the ways in which the most *relevant* and valued composition come into being.

«This is the telepathic imperative. Data exists everywhere in excess. In the wireless city, it floats and settles in a hazy surround. Sifting through the modalities of dust to sense and communicate through the urban medium will ultimately require a well tuned telepathic sense.»[30]

In his *Amusing Ourselves to Death*[31], Neil Postman hypothizes how the realization of these complex media ecologies would expose us to this type of issue: for the quantity and quality of information; for their structural configuration (in the sense of the type of media and, within it, of the architecture of information); for their shape (this hypothesis was even more strongly confirmed in *Informing Ourselves to Death*). The problem, according to Postman, is not in the availability of information, but in the possibility to extract meaning from information.

This type of problem has been highlighted multiple times, and defined as *information overload*, *data smog*, *spam*, or under the constructivist form of the *attention economy* described by Davenport and Beck[32]. The technological solutions at this level are also problematic, at least when they are not oriented towards providing usable, accessible and inclusive mechanisms for content classification, filter and for the expression of their relevance. And – also in these latter cases –, the algorithmic dimensions of these processes isolate us from the possibility to comprehend the meaning of information, however remixed. Google's *Filter Bubble*[33] is a classic example of this phenomenon, with enormous impacts on the reachability of information, on its accessibility, on the rights to expression and information, and on the dangers deriving from the creation of opaque zones in which the mechanisms according to which information are published, hidden or highlighted are not transparent.

Technical solution apart – and their corresponding algorithms, systems, interfaces, constantly more advanced to be able to enormous amounts of data, information and knowledge – the most interesting results come from the *transmedia* character of information, and from their *participatory performability*.

From the first point of view, following Jenkins[34]' definitions, transmediality allows us to simplify the extraction of meaning from enormous amounts of information, and making its access more immersive and accessible: content which is sharable; spreadable; which offer opportunities for mutual interconnection, across different media.

---

29  Gabrys, Jennifer. "Telepathically Urban." In Circulation and the City: Essays on Urban Culture, edited by Alexandra Boutros and Will Straw, 48–63. Montreal: McGill-Queen's University Press, 2010.
30  ibid.
31  Postman, Neil. Amusing Ourselves to Death: Public Discourse in the Age of Show Business. New York: Penguin, 1985.
32  Davenport, Thomas H., and John C. Beck. The Attention Economy: Understanding the New Currency of Business. Cambridge: Harvard Business School Press, 2001.
33  The concept of Filter Bubble, initiated by activist Eli Parser, indicates those algorithms which, by highlighting certain search results – for example on Google – on the basis of the interpretation of what could be more pertinent to us according to our user profile – and on how this is interpreted by online systems –, effectively block us from the possibility to access all of the available information, creating, thus, a bubble around us, and avoind us from leaving it. For more information: http://en.wikipedia.org/wiki/Filter_bubble
34  Jenkins, Henry. Convergence Culture: Where Old and New Media Collide. New York: NYU Press, 2006.

From the second point of vies – which becomes important also evaluating the first one –, the problem of *overload* and of the impossibility to extract meaning becomes easier to confront to when messages are freely accessible and performable, and when the ways in which they have been generated is transparent and also accessible, just as the way in which it should be possible to intervene in the flows of of their generation, processing – and remix/recombination –, and their propagation.

In synthesis, this equals to the need to create legibility for the relational graphs related to the generation, processing and propagation of data, information and knowledge, and to make accessible – in ways which are inclusive – the possibility to intervene, infiltrate and add in any stage of the process: enabling information to be *performable*.

Both mechanisms require intellectual property management techniques which are more refined, advanced and just, if compared to the ones we have available now, from legal and perceptive points of view.

In the discussion about the Planetary Garden[35], Clément proposes specific questions regarding property which are of fundamental importance in all of this discussion. Ecological dynamics assume the restructuring of the applicability of private property, from the point of view of a mutation of the concept of *value,* and from the point of view of the emergence of what can be described as the *dreaming economist,* guarantor of a dynamic, mutating and mutant landscape, not a definitive one.

«Emergent ecosystems could be a source of wealth, but being misunderstood by the system, they are misunderstood by us, as well.»[36]

These dynamics attribute a central role to knowledge and to its free accessibility, recombination, remix, both in terms of usage and in the ones of imagination, education and sharing.

The redefinition processes for the concept of property (intellectual, in this case) – and on its implications on accessibility, inclusiveness and usability – become necessary when objects themselves undergo radical transformation.

Complex mutations have already happened to be able to confront with entire market disruptions brought on by the diffusion of mp3, videos, images and other simple media (simple, in the sense of mono-media). Thus, it can become intuitive how even more radical transitions and transformations will be needed to adapt these concepts to data, information and knowledge which, now, are of a completely different type.

Services like Google, Facebook, Twitter, Amazon, Apple, produce data, information, knowledge, identities which are diffused across different and multiple devices and modalities, interacting in profound ways with the things we know – and that we can know – about the world, its inhabitants, and with the ways in which we experience places, events, monuments, schools, restaurants, workplaces and a lot more.

Messaging applications reach us ubiquitously.

---

35  For example on Domus Magazine: http://www.domusweb.it/it/recensioni/2013/11/08/lezione_di_giardino.html
36      *ibid*

Devices, sensors, gadgets, wearable technologies, prosthetics and, soon, entire body and neural extensions interconnect bodies, emotions, health information, movements, gestures, sensations, exhibiting them on social networks and sharing them – knowing or unknowing, whether we like it or not – with diverse types of services and processes, with human beings and machines.

Furthermore, algorithms create additional dimensions, in which each gesture, movement or action can be recombined with others, and transformed into information and knowledge.

These are territories for which there are no maps, yet. We find ourselves within a grey area in which laws, regulations and people's perceptions are not defined.

In this scenario it could be desirable to actuate a shared, open and inclusive effort to define the Ubiquitous Commons[37], the commons in the era of ubiquitous technologies.

---

37  For more information regarding the Ubiquitous Commons: http://www.ubiquitouscommons.org/